\documentstyle[12pt,aaspp4,epsf]{article} 
\begin{document} 

\title{Hard X-ray Luminosities of Multinuclei Infrared 
Luminous Galaxies Showing a Radio/Far-Infrared Excess}
\author{Masatoshi Imanishi \altaffilmark{1}}  
\affil{National Astronomical Observatory, Mitaka, Tokyo 181-8588, Japan}
\author{and \\
Shiro Ueno \altaffilmark{2}} 
\affil{Department of Physics and Astronomy, University of Leicester,  
University Road, Leicester LE1 7RH, UK}
 
\altaffiltext{1}{Present address: Institute for Astronomy, 
University of Hawaii, 2680 Woodlawn Drive, Honolulu, Hawaii 96822, 
USA}

\altaffiltext{2}{Present address:  
Space Utilization Research Program, Tsukuba Space Center, 
National Space Development Agency of Japan,  
2-1-1 Sengen, 
Tsukuba 305-8505, Japan}

\begin{abstract} 
 
We report the results of hard X-ray observations of four 
multinuclei merging infrared luminous galaxies (IRLGs).
We selected these four sources for their excess of radio to 
far-infrared luminosity ratio compared with starburst galaxies.
This excess suggests that activity associated with 
a supermassive black hole (SMBH) contributes strongly to the 
IRLGs' bolometric luminosities.
Although we expect strong hard X-ray emission from the SMBH-driven 
activity, the radio-excess multinuclei merging IRLGs show considerably 
smaller hard X-ray luminosities relative to far-infrared 
(40$-$500 $\mu$m) and infrared (8$-$1000 $\mu$m)   
luminosities than active galactic nuclei (AGNs) 
showing a similar radio-excess.
This result may demonstrate that emission in the hard X-ray region 
from SMBH-driven activity in the multinuclei merging IRLGs is 
severely suppressed compared to a typical spectral energy distribution 
of SMBH-driven activity in AGNs.
If this is a common property of merging IRLGs, 
without its correction, hard X-ray observations 
underestimate the contribution of SMBH-driven activity to the 
bolometric luminosities of merging IRLGs.

\end{abstract} 
 
 
\section{Introduction} 

Infrared luminous galaxies (IRLGs) radiate most of their extreme 
luminosities in the far-infrared 
(FIR; 40$-$500 $\mu$m; 
{\it L}$_{\rm FIR}$ $>$ 10$^{11.25}${\it L}$_{\odot}$; 
{\it H}$_{0}$ $=$ 75 km s$^{-1}$ Mpc$^{-1}$, {\it q}$_{0}$ $=$ 0.5). 
Most of these sources are thought to be in some stage of a merger 
process of molecular gas-rich galaxies and to evolve into 
active galactic nuclei (AGNs) after the merger 
(Sanders et al. 1988a; Hutchings \& Neff 1991).
In IRLGs, two types of activity are thought to occur:  
one associated with a supermassive black hole 
(SMBH) and the other, with a starburst 
(e.g., Sanders \& Mirabel 1996).
Which of the two types of activity, starburst or SMBH, is the dominant 
energy source is a fundamental issue in the study of IRLGs.

Since the UV to soft X-ray ionizing photons are the main power source 
of SMBH-driven activity, 
measuring the UV to soft X-ray luminosities of SMBH-driven 
activity is the most direct way to estimate the contribution 
of SMBH-driven activity to the bolometric luminosities of galaxies.
However, direct measurements of the UV to soft X-ray luminosity 
of SMBH-driven activity is practically impossible 
in IRLGs because of high attenuation by the abundant dust in IRLGs.
Therefore, other less-direct methods have been employed to estimate 
the contribution of SMBH-driven activity to the bolometric luminosities.

One such method is hard X-ray observations; following convention, 
we refer to the 2$-$10 keV energy band as hard X-rays throughout 
this paper.
Since the effect of extinction is small in the hard X-ray region, 
SMBH-driven activities should be detectable even through thick 
obscuring material, up to {\it N}$_{\rm H}$ $\sim$ 10$^{24}$ cm$^{-2}$.
Further, the hard X-ray luminosities relative to far-infrared 
luminosities ({\it L}$_{\rm X}$/{\it L}$_{\rm FIR}$) in AGNs 
(${\ ^{\displaystyle >}_{\displaystyle \sim}\ }$0.1) are 
2$-$3 orders of magnitude higher than those in starburst galaxies 
($\sim$10$^{-3}$; Tsuru 1992).
The strong hard X-ray emission from AGNs is thought to originate from 
SMBH-driven activity.
If SMBH-driven activity in IRLGs has the same hard X-ray emission 
efficiency as that in AGNs, 
the main energy source of IRLGs should be easily discernible 
from the intrinsic hard X-ray luminosities. 
Hard X-ray observations of IRLGs have been made 
and no strong hard X-ray emission  has been found 
(i.e., {\it L}$_{\rm X}$/{\it L}$_{\rm FIR}$ $<$ 0.05; Ueno et al. 1996;  
Nakagawa et al. 1997; Ogasaka et al. 1997; Brandt et al. 1997; 
Iwasawa 1999).
These results can be explained if the IRLGs are powered by starburst 
activity rather than SMBH-driven activity, or if the direct hard X-ray 
emission from SMBH-driven activity is completely blocked 
by thick ($N_{\rm H}$ $>$ 10$^{24}$ cm$^{-2}$) absorbing gas.
However, many IRLGs are thought to be powered by SMBH-driven 
activity based on near-infrared  (Veilleux et al. 1997b) and 
mid-infrared diagnostics (Genzel et al. 1998), 
both of which suggest huge UV to soft X-ray emission from SMBH-driven 
activity in these IRLGs.
Some of such IRLGs have been observed  
with hard X-rays (e.g., IRAS F20460+1925; Ogasaka et al. 1997;  
IRAS F23060+0505; Brandt et al. 1997; Mrk 231; Iwasawa 1999;   
Mrk 463; Ueno et al. 1996).
Although direct hard X-ray emission is thought to be 
detected based on the spectral shape, 
the extinction-corrected hard X-ray luminosities are much smaller than 
that expected if they are powered by SMBH-driven activity.
This may indicate that the intrinsic hard X-ray luminosities 
relative to UV to soft X-ray luminosities from 
SMBH-driven activities in IRLGs are considerably smaller than those 
in AGNs.
In other words, even if strong UV to soft X-ray emission from 
SMBH-driven activity exists, hard X-ray emission from SMBH-driven 
activity may not develop during the early phase of a merger as 
much as during the later or final stages of a merger 
(Imanishi \& Ueno 1999).

To investigate the possible hard X-ray underluminosity 
(relative to other spectral regions) of SMBH-driven activity during 
a merger, we performed hard X-ray observations of merging IRLGs 
that are thought to possess strong SMBH-driven activities.

\section{Targets}

To select merging IRLGs possessing strong SMBH-driven activity, 
we use the radio to far-infrared luminosity ratio.
In galaxies where starburst activity is thought to be a predominant 
energy source, the distribution of radio 1.5 GHz (= 20 cm) to 
far-infrared luminosity ratios is very narrow.
If we use the {\it q}-value introduced by 
Helou, Soifer, \& Rowan-Robinson (1985), 
\begin{center}
{\it q}  $\equiv$  log({\it f}$_{\rm FIR}$/{\it f}$_{\rm 20cm}$)\\
{\it f}$_{\rm FIR}$ =  0.336 $\times$ (2.58 $\times$ 
{\it f}$_{\rm 60\mu m}$+{\it f}$_{\rm 100\mu m}$),\\
\end{center}
where {\it f}$_{x}$ means flux at wavelengths ``x'' in units of Jy, 
{\it q}-values of starburst galaxies are {\it q} = 2.34 $\pm$ 0.19 
at {\it L}$_{\rm FIR}$ $=$ 10$^{9}$ $-$ 10$^{12}${\it L}$_{\odot}$ 
(Condon et al. 1990, 1991a).
If, besides starburst activity, SMBH-driven activity 
(both radio-loud and -quiet cases) 
contributes strongly to the bolometric luminosity of a galaxy, 
strong radio synchrotron emission is expected to exist.
Since this radio emission is presumably not directly 
coupled with far-infrared emission, the {\it q}-value is expected 
to be decreased.
Actually, not only radio-loud AGNs but also many radio-quiet 
Seyfert galaxies show {\it q}-values smaller than those of 
starburst galaxies 
(Knapp et al. 1990; Elvis et al. 1994; Dahari \& De Robertis 1988). 
Given that the rest of radio-quiet Seyfert galaxies show {\it q}-values 
similar to or larger than those of starburst galaxies, 
the {\it q}-value may not be a very sensitive indicator for the presence 
of SMBH-driven activity, that is, the decrease of the {\it q}-values 
may be recognized only when the contribution of SMBH-driven activity 
relative to starburst activity is high.
Therefore, if the {\it q}-value of a galaxy is clearly smaller than 
those of starburst galaxies, we can claim with some assurance that 
the galaxy is strongly powered by SMBH-driven activity.

At the highest far-infrared luminosity range, the mean of q-value 
increases slightly, and no starburst galaxies with 
{\it L}$_{\rm FIR}$ $>$ 10$^{11.25}${\it L}$_{\odot}$ 
show {\it q}-values smaller than 2.05 
(Condon et al. 1991b; we plot the q-value distribution of starburst 
galaxies with {\it L}$_{\rm FIR}$ $>$ 10$^{11.25}${\it L}$_{\odot}$ 
in the upper panel of Fig. 3).
We selected four multinuclei merging IRLGs with {\it q} $<$ 2.0 
as candidates strongly powered by SMBH-driven activities.
 
The selected targets are PKS 1345+12 (= IRAS 13451+1232, 4C 12.50),  
Super-Antennae (= IRAS 19254$-$7245), IRAS 04154+1755, 
and Mrk 1224 (= IRAS 09018+1447, UGC 4756). 
Properties of these galaxies are summarized in Table 1.
 
PKS 1345+12 is a merging double-nuclei IRLG 
(Murphy et al. 1996; Surace \& Sanders 1999). 
The {\it q}-value is $-$0.35 (Kim 1995).
The optical spectrum is a Seyfert 2 type (Sanders et al. 1988b).
Veilleux et al. (1997b) detected a broad (FWHM $\sim$ 2500 km s$^{-1}$) 
near-infrared Pa$\alpha$ emission 
line, whose intrinsic luminosity is as high as optically selected 
quasars.

Super-Antennae is a merging double-nuclei IRLG 
(Mirabel, Lutz, \& Maza 1991; Duc, Mirabel, \& Maza 1997).
It shows an excess of radio 
(4.9 GHz = 6 cm) to far-infrared luminosity ratio compared to 
starburst galaxies (Roy \& Norris 1997).
Assuming a radio spectrum of $\nu^{-1}$, we find a {\it q} = 1.31. 
The optical spectrum is a Seyfert 2 type 
(Mirabel et al. 1991; Duc et al. 1997).
Mid-infrared observation of the 7.7 $\mu$m 
polycyclic aromatic hydrocarbon emission to 
7.7 $\mu$m continuum luminosity ratio suggested that SMBH-driven 
activity contributes strongly to the bolometric luminosity 
(Genzel et al. 1998).

IRAS 04154+1755 has a {\it q}-value of 1.93, a Seyfert 2 type optical 
spectrum, and a one-sided radio jet structure (Crawford et al. 1996).
In the Digitized Sky Survey image, IRAS 04154+1755 seems to be 
a merging system of two galaxies.
We obtained near-infrared J- and K$'$-band images of 
IRAS 04154+1755 using QUIST at the University of Hawaii 0.6m telescope 
(Hodapp, Hora, \& Metzger 1997).
The image also shows two nuclei that may be interacting each other.

Mrk 1224 is a merging multinuclei IRLG (Smith et al. 1996).
The {\it q}-value is 1.97, and two radio sources, which 
correspond to the two brightest nuclei of these galaxies, are seen  
(Crawford et al. 1996).
The optical spectral classification is unknown.

\section{Observation and Data Analysis}

PKS 1345+12, Super-Antennae, IRAS 04154+1755, and Mrk 1224 were 
observed with the Advanced Satellite for Cosmology and Astrophysics 
({\it ASCA}; Tanaka, Inoue, \& Holt 1994).
The observing log is summarized in Table 2.
The observations were performed with the two solid-state imaging 
spectrometers (SIS0 and SIS1) and the two gas imaging spectrometers 
(GIS2 and GIS3).
We removed data obtained under high-background environments, 
and used only high- and medium-bit data.
The remaining effective exposure time after this data selection 
and net source counts are also summarized in Table 2.

Standard calibration and data reduction techniques were employed 
using FTOOLS software provided by the {\it ASCA} Guest Observer Facility.
The separation of the nuclei for these sources is 
in the range 2$''-$20$''$, 
which is far smaller than the spatial resolution of {\it ASCA} 
(2$'$ for 50\% encircled energy).
Hence, the hard X-ray luminosities derived from {\it ASCA} are 
the sum of the emission from the nuclei of each source. 

Significant X-ray photons were detected at the position of PKS 1345+12 
and Super-Antennae in the SIS and GIS detectors.
We extracted X-ray spectra from circular regions centered on the 
source with radii 2$\farcm$6 and 3$\farcm$3 for the SIS and GIS data of 
PKS 1345+12, and 2$\farcm$1 and 3$\farcm$4 for the SIS and GIS data of 
Super-Antennae, respectively.
We extracted background spectra from an annular region outside 
the source region.

For IRAS 04154+1755 data, we marginally detected a source 
only in the SIS1 image.
Because of the position coincidence with the radio position 
(Crawford et al. 1996) within the 
absolute pointing accuracy of {\it ASCA} ($\sim$1$'$), we conclude 
this source is IRAS 04154+1755.
We extracted X-ray spectra from circles of radius 2$\farcm$1 
centered on the source.
For detectors other than the SIS1, we found no source candidates.
For these detectors, we estimated the flux around the coordinate 
where a source was detected in the SIS1 image.
We accumulated the X-ray photons with radii of 2$\farcm$1 and 
3$\farcm$4 for the SIS0 and GIS detectors, respectively.
Background subtraction was done in the same way as before.

For Mrk 1224 data, we identified a source candidate in the SIS0 and 
SIS1 images.
Because of the position coincidence with the radio position 
(Crawford et al. 1996) within the 
absolute pointing accuracy of {\it ASCA}, we conclude 
this source is Mrk 1224.
We extracted X-ray spectra from circles of radius 1$\farcm$6 centered on 
the source.
For the GIS detectors, we found no source candidates.
For them, we estimated the flux around the coordinate where a source was 
detected in the SIS images.
We accumulated the X-ray photons with a radius 3$\farcm$4 for the GIS 
detectors.
Background subtraction was done in the same way as before.

\section{Results}

\subsection{PKS 1345+12}

The SIS and GIS spectra of PKS 1345+12 are shown in 
Figures 1a and 1b, respectively.
For clarity, we show the SIS and GIS data separately.  

The Galactic absorption toward this source is interpolated to be 
{\it N}$_{\rm H}$ = 1.9 $\times$ 10$^{20}$ cm$^{-2}$ from 
the Einstein On-Line Service (EINLINE; 131.142.11.73).
Since this absorption is negligible for 2$-$10 keV band, 
we neglect it for the following spectral fitting.
We first fit the data with a single power-law model modified by 
cold absorption at the rest frame (z = 0.121).
We fit the combined SIS and GIS spectra simultaneously.
The fitting results are summarized in Table 3, and are plotted 
in Figures 1a and 1b.
The extinction-corrected hard X-ray luminosity is 
{\it L}$_{\rm X}$ = 3.2 $\times$ 10$^{43}$ ergs s$^{-1}$, 
which corresponds to 
{\it L}$_{\rm X}$/{\it L}$_{\rm FIR}$ = 8.5 $\times$ 
10$^{-3}$ (= 10$^{-2.07}$).

The derived photon index ($\Gamma$ = 0.82$^{+0.07}_{-0.03}$) is smaller 
than that for typical AGNs ($\Gamma$ $\sim$ 1.7).
This might be intrinsic, but such a small $\Gamma$ can be reproduced 
if we fit poor S/N data whose intrinsic spectra are a highly absorbed 
power-law of $\Gamma$ = 1.7.
For comparison, we fix $\Gamma$ as 1.7 and fit with 
single power-law model modified by cold absorption at the rest frame.
The fitted parameters are summarized in Table 3.
This model is also acceptable in terms of reduced $\chi^{2}$ value.
The extinction-corrected hard X-ray luminosity is consistent 
with the above result within 10\%.
We adopt the value of {\it L}$_{\rm X}$ = 3.2 $\times$ 10$^{43}$ 
ergs s$^{-1}$ for the discussion.

No significant iron K$\alpha$ emission line is detected near 
5.7 keV (= 6.4 keV/1.121), with the equivalent width 
of $<$466 eV at the 90\% confidence level.
This value is smaller than 
those whose direct hard X-ray emission is completely blocked 
and for which only a scattered component is seen (equivalent width 
$>$1 keV; 
Koyama et al. 1989; Iwasawa et al. 1993; Iwasawa \& Comastri 1998; 
Maiolino et al. 1998; Matt et al. 1996), suggesting that the observed 
hard X-ray emission is not a scattered component. 

The detection of broad near-infrared Pa$\alpha$ emission 
toward this source by Veilleux et al. (1997b) suggests 
this is a less obscured system by dust.
Hard X-ray absorption is caused mainly by gas, while near-infrared  
extinction is caused by dust.
If the gas-to-dust ratio toward this source was far higher than usual 
Galactic value ({\it N}$_{\rm H}$/$A_{V}$ $=$ 1.8 $\times$ 10$^{21}$ 
mag$^{-1}$ cm$^{-2}$; Predehl \& Schmitt 1995), 
direct hard X-ray emission could be completely blocked even toward 
a source less obscured by dust.
However, the hard X-ray spectra of sources with detected broad 
near-infrared lines (NGC 2992, MCG $-$05-23-16, IRAS F20460+1925, 
IRAS F23060+0505; 
Veilleux et al. 1997a, b), all of which are of higher quality than 
that of PKS 1345+12, show small X-ray absorption  
({\it N}$_{\rm H}$ $<$ 10$^{23}$ cm$^{-2}$; Turner et al. 1997; 
Ogasaka et al. 1997; Brandt et al. 1997). 
Therefore, we believe we see direct hard X-ray emission from 
SMBH-driven activity in PKS 1345+12.
 
\subsection{Super-Antennae}

The SIS and GIS spectra of 
Super-Antennae are shown in Figures 2a and 2b, respectively.

We fit the data with a single power-law model modified by 
cold absorption at the rest frame (z = 0.062).
We fit the combined SIS and GIS spectra simultaneously.
Since the expected {\it N}$_{\rm H}$ at the Galactic latitude of 
this source ($-$29$^{\circ}$) is negligible for 2$-$10 keV band 
({\it N}$_{\rm H}$ = 0$-$2 $\times$ 10$^{21}$ cm$^{-2}$; 
Dickey and Lockman 1990), 
we do not include the Galactic absorption for the spectral fitting.
The fitting results are summarized in Table 3, and are plotted 
in Figures 2a and 2b.
The extinction-corrected hard X-ray luminosity is 
{\it L}$_{\rm X}$ = 1.4 $\times$ 10$^{42}$ ergs s$^{-1}$, 
which corresponds to 
{\it L}$_{\rm X}$/{\it L}$_{\rm FIR}$ = 5.3 $\times$ 10$^{-4}$ 
(= 10$^{-3.28}$).

The derived photon index ($\Gamma$ = 1.1$^{+0.1}_{-0.2}$) is smaller 
than that for typical AGNs ($\Gamma$ $\sim$ 1.7).
For comparison, we fix $\Gamma$ as 1.7 and fit with 
single power-law model modified by cold absorption at the rest frame, 
as described in $\S$ 4.1.
The fitted parameters are summarized in Table 3.
This model is also acceptable in terms of reduced $\chi^{2}$ value.
The extinction-corrected hard X-ray luminosity is consistent 
with the above result within 20\%.

The equivalent width of the iron K$\alpha$ emission line at 6.0 keV 
(= 6.4 keV/1.062) is $<$1630 eV at the 90\% confidence level.
Because the upper limit of the equivalent width is large, we cannot 
rule out the possibility that the detected hard X-ray emission is a 
scattered component.
If this is the case, the intrinsic hard X-ray luminosity could be much 
higher than the above value, and we must know the scattering efficiency 
to estimate intrinsic hard X-ray luminosity. 
To estimate the scattering efficiency, 
we must detect direct emission from SMBH-driven activities, which could 
be visible at higher energy ($>$10 keV).
The hard X-ray scattering efficiency ({\it f}$_{\rm scatt}$) 
is estimated to be $\sim$3\% for the nearby Seyfert 2 galaxy, NGC 4945, 
which is the only object known so far whose 
2$-$10 keV hard X-ray emission is completely blocked 
({\it N}$_{\rm H}$ $>$ 10$^{24}$ cm$^{-2}$) 
and whose scattering efficiency is estimated based on the direct 
emission detected at $>$10 keV (Iwasawa et al. 1993).
Though its far-infrared luminosity 
({\it L}$_{\rm FIR}$ = 10$^{10.74}$ {\it L}$_{\odot}$) 
is not as high as IRLGs 
({\it L}$_{\rm FIR}$ $>$ 10$^{11.25}$ {\it L}$_{\odot}$), 
intense far-infrared emission from the compact, dusty central region 
of the galaxy (Rice et al. 1988) and the coexistence of SMBH-driven 
and starburst activities 
(Moorwood \& Oliva 1994) suggest that NGC 4945 has similar nuclear 
environment to merging IRLGs.
We tentatively adopt the value of 3\% as a representative value 
of hard X-ray scattering efficiency in merging IRLGs.

If we assume a power-law spectrum of $\Gamma$ = 1.7 and a scattering 
efficiency of 3\%, then the intrinsic hard X-ray luminosity is 
2.9 $\times$ 10$^{43}$ $\times$ (0.03/{\it f}$_{\rm scatt}$) 
ergs s$^{-1}$, 
or {\it L}$_{\rm X}$/{\it L}$_{\rm FIR}$ $=$ 1.1 $\times$ 10$^{-2}$ 
(= 10$^{-1.97}$) $\times$ (0.03/{\it f}$_{\rm scatt}$).

\subsection{IRAS 04154+1755}

A source was detected in the SIS1 detector at a 4.8 $\sigma$ level, 
but not in the other detectors ($<$2 $\sigma$).  
Hence, it is impossible to obtain any useful spectral information.  
By neglecting the small Galactic absorption 
({\it N}$_{\rm H}$ = 1.8 $\times$ 10$^{21}$ cm$^{-2}$; EINLINE), 
we estimate the upper limit 
of the hard X-ray flux at the 90\% confidence level by using 
the combined SIS and GIS data.

We assume 
(1) a power-law spectrum of $\Gamma$ = 1.0, as observed 
for PKS 1345+12 and Super-Antennae, 
(2) a power-law spectrum of $\Gamma$ = 1.7 modified 
by cold absorption with {\it N}$_{\rm H}$ = 10$^{23}$ cm$^{-2}$ 
at the rest frame (z = 0.056), 
and 
(3) that direct hard X-ray emission is completely blocked 
    ({\it N}$_{\rm H}$ $>$ 10$^{24}$ cm$^{-2}$) and 
    only a scattered component ($\Gamma$ = 1.7) is seen 
    with a scattering efficiency of 3\%.
The estimated hard X-ray flux and extinction-corrected hard X-ray 
luminosity are summarized in Table 3.

\subsection{Mrk 1224}

A source was detected in the SIS detectors, but the detection was 
only $\sim$5 $\sigma$ (SIS0) and $\sim$3 $\sigma$ (SIS1).
For the GIS detectors, source detection was $<$3 $\sigma$ for each.
It is therefore impossible to obtain any useful spectral information.  
We neglect the small Galactic absorption 
({\it N}$_{\rm H}$ $=$ 3.7 $\times$ 10$^{20}$ cm$^{-2}$; EINLINE), and 
estimate the upper limit of the hard X-ray flux 
and extinction-corrected hard X-ray luminosity at the 90\% 
confidence level by using the combined SIS and GIS data, 
as we did with the IRAS 04154+1755 data.
The results are summarized in Table 3.

\section{Discussion} 

\subsection{Hard X-ray Luminosity Relative to Far-Infrared and 
Infrared Luminosity}

In Table 4, we summarize the hard X-ray emission properties of the four 
radio-excess multinuclei merging IRLGs.
For comparison, we examine the hard X-ray emission properties 
of AGNs (= Seyfert type optical spectra) 
and radio-excess multinuclei merging IRLGs in the literature. 
We summarize the results in Table 5.
To find hard X-ray data, we first examined Turner et al. (1997) 
and Nandra et al. (1997) for {\it ASCA} data, 
Nandra \& Pounds (1994) for {\it Ginga} data, 
and Elvis et al. (1994) for {\it HEAO1} data.
We include additional sources for which hard X-ray data 
are available in other papers.
If a source was observed both with {\it Ginga} and {\it ASCA}, 
we adopt the {\it ASCA} data, but show the {\it Ginga} data for 
comparison in Table 5.
This is because time variation of the hard X-ray luminosity from 
SMBH-driven activity is sometimes significant.
In our sample, NGC 2992 and Mrk 3 show time variations of 
a factor of $>$4.
We then examined the far-infrared and radio 1.5 GHz data for these AGNs.
If radio data were taken at different epochs, we adopted the latest data, 
though the differences were always less than a factor of 1.8, 
corresponding to $\Delta${\it q} $<$ 0.25.
In some cases, we estimated radio 1.5 GHz flux from measurements  
at other radio frequencies near 1.5 GHz (1.4, 2.3, and 4.8 GHz).
In Table 5, we compile all AGNs for which an intrinsic hard X-ray 
luminosity has been derived, and far-infrared and radio data are 
available.
However, we do not include the objects whose hard X-ray emission is 
convincingly a scattered component based both on a large equivalent width 
of the iron K$\alpha$ emission line and on the spectral shape 
in the hard X-ray band, where the intrinsic hard X-ray luminosity is 
highly uncertain. 
These are NGC 1068 (Koyama et al. 1989), NGC 4388 
(Netzer, Turner, \& George 1998), NGC 7674 (Malaguti et al. 1998), and 
NGC 6240 (Nakagawa et al. 1997; Iwasawa \& Comastri 1998).
We do not include flat spectrum compact-radio sources in Elvis et al. 
(1994; 3C273, Q0007+106, Q1028+313, Q1721+343), because their hard X-ray 
emission could be relativistically beamed toward our line of sight.

In Figure 3, we plot {\it L}$_{\rm X}$/{\it L}$_{\rm FIR}$ 
ratios of multinuclei merging IRLGs and AGNs against 
observed {\it q}-values, where we add Mrk 463 to our sample of 
radio-excess multinuclei merging IRLGs.
For radio-excess sources ({\it q} $<$ 2.0), which we regard as sources 
strongly powered by SMBH-driven activity 
%------------------------
\footnote{
We note a galaxy can be powered by starburst activity even though 
the galaxy shows a Seyfert type optical spectrum 
(e.g., Genzel et al. 1995).
For {\it q} $>$ 2.0 AGNs, we have no strong evidence that they 
are powered by SMBH-activity.
Actually, the {\it L}$_{\rm X}$/{\it L}$_{\rm FIR}$ ratios 
distribute from 10$^{-4}$ to 10$^{0}$ in Figure 3.
}, 
%------------------------
we find a clear difference of {\it L}$_{\rm X}$/{\it L}$_{\rm FIR}$ 
ratios between AGNs and multinuclei merging IRLGs.
The {\it L}$_{\rm X}$/{\it L}$_{\rm FIR}$ of radio-excess AGNs is 
actually high as we anticipated, but 
the five radio-excess multinuclei merging IRLGs 
(the four merging IRLGs and Mrk 463) show 
{\it L}$_{\rm X}$/{\it L}$_{\rm FIR}$ ratios far smaller than 
the AGNs with a similar radio-excess.
The AGN NGC 2992 shows a small $L_{\rm X}$/$L_{\rm FIR}$ 
($\sim$10$^{-2.12}$) ratio, but the hard X-ray luminosity derived from 
{\it ASCA}, 
which we adopted, is about 4 times smaller than that derived from 
{\it Ginga} data.
Hence, the {\it ASCA} observation of NGC 2992 may have been made 
during a very faint phase.
All the radio-excess AGNs in Table 5 show single-nucleus morphology, 
and an $IRAS$ 25 $\mu$m to 60 $\mu$m flux ratio of 
${\ ^{\displaystyle >}_{\displaystyle \sim}\ }$0.2 
(i.e., ``warm color''; Sanders et al. 1988b), indicating they are 
in the later or final stage of merging.
This suggests that, during the early stage of a merger, 
the $L_{\rm X}$/$L_{\rm FIR}$ ratio appears to be considerably smaller 
than that in the later or final stage, 
even though a similar radio-excess is found.

Most of multinuclei merging IRLGs show a spectral energy distribution 
that peaks at 60 $\mu$m and radiates a predominant fraction of infrared  
(8$-$ 1000$\mu$m) emission in the far-infrared (40$-$500 $\mu$m), 
while many ordinary AGNs do not.
This different spectral energy distribution in the infrared region 
could cause the difference in the $L_{\rm X}$/$L_{\rm FIR}$ ratios 
in Figure 3.
We compare the extinction-corrected hard X-ray luminosities with 
infrared luminosities ({\it L}$_{\rm IR}$).
The {\it L}$_{\rm X}$/{\it L}$_{\rm IR}$ ratios 
plotted against observed {\it q}$'$-values ($\equiv$ 
log {\it f}$_{\rm IR}$/{\it f}$_{\rm 20 cm}$) 
are shown in Figure 4.
The radio-excess multinuclei merging IRLGs still show 
$L_{\rm X}$/$L_{\rm IR}$ ratios smaller by a large factor than 
those of AGNs with a similar radio-excess.

\subsection{Possible Explanations and Implications}

We consider four possible explanations for the small 
$L_{\rm X}$/$L_{\rm FIR}$ and $L_{\rm X}$/$L_{\rm IR}$ ratios 
in radio-excess multinuclei merging IRLGs compared with radio-excess 
AGNs. 
(1) A high fraction of UV to soft X-ray emission is absorbed by dust 
in the multinuclei merging IRLGs, 
(2) direct hard X-ray emission suffers partial absorption 
in the multinuclei merging IRLGs, 
(3) direct hard X-ray emission is completely blocked, and 
nearly no scattered component is seen in the multinuclei merging IRLGs, 
and 
(4) emission in the hard X-ray region from SMBH-driven activity in 
the multinuclei merging IRLGs is severely suppressed compared to 
a typical spectral energy distribution of SMBH-driven activity in AGNs.
We will consider each explanation below and suggest that the fourth one 
is the most plausible to us.

Let us consider the first explanation.
If the fraction of UV to soft X-ray emission from SMBH-driven activity 
absorbed by dust and reradiated in the infrared  is high 
in the multinuclei merging IRLGs, the $L_{\rm IR}$ will increase, 
and thus $L_{\rm X}$/$L_{\rm IR}$ will become apparently smaller.
Since a mass concentration of molecular gas toward the nuclei 
is found in some IRLGs 
(Scoville, Yun \& Bryant 1997, Bryant \& Scoville 1996), 
the high fraction is quite possible.
However, the fraction of UV to soft X-ray emission absorbed by dust 
and reradiated in the infrared region in AGNs 
is $\sim$30$-$50\% (Sanders et al. 1989; Rowan-Robinson 1995).
Even if we consider nearly 100\% of UV to soft X-ray emission from 
SMBH-driven activity is absorbed by dust in the multinuclei merging 
IRLGs, 
the {\it L}$_{\rm X}$/{\it L}$_{\rm IR}$ could decrease 
by only a factor of 2$-$3.
Hence, this explanation alone cannot explain the observed large 
difference.

Next, we consider the second explanation that the hard X-ray emission 
suffers partial absorption by X-ray absorbing gas such that 
a large fraction of the X-ray emission is completely blocked 
but the remaining small fraction is relatively transparent.
If this is the case, we cannot estimate correctly the hard X-ray 
extinction from the observed hard X-ray spectra, and thus the intrinsic 
hard X-ray luminosity could be higher than the derived 
extinction-corrected hard X-ray luminosity from the observed spectra.
However, the intrinsic size of X-ray emission from SMBH-driven 
activity is considered very compact ($<$10$^{-3}$ pc; 
Barr \& Mushotzky 1986).
To achieve partial absorption, the size of the clumps of 
the absorbing gas must be far smaller than 10$^{-3}$ pc, and 
yet each clump must have column density 
higher than {\it N}$_{\rm H}$ $>$ 10$^{24}$ cm$^{-2}$.

We then consider the third explanation that direct hard X-ray 
emission is completely blocked and nearly no scattered component is seen.
However, the detection of near-infrared broad emission lines 
toward PKS 1345+12 and Mrk 463 (Veilleux et al. 1997a,b) 
indicates we detect direct hard X-ray emission from SMBH-driven activity 
in these sources, as discussed in $\S$ 4.1.
For other three sources, Super-Antennae, IRAS 04154+1755, and Mrk 1224, 
the detection of broad optical or near-infrared emission lines has 
not been reported.
If a direct component is completely blocked and nearly no scattered 
component is seen ({\it f}$_{\rm scatt}$ $<$ 0.5\%), 
intrinsic {\it L}$_{\rm X}$/{\it L}$_{\rm IR}$ ratios could be 
in the range of AGNs.
Super-Antennae and IRAS 04154+1755 show Seyfert 2 type optical spectra.
This suggests a large amount of 
ionizing UV to soft X-ray photons from SMBH-driven activity are 
escaping in directions roughly perpendicular to our line of sight 
(i.e., the SMBH-driven activity is {\it not} completely 
obscured toward {\it all} directions).
We expect the scattering efficiency is high in the multinuclei 
merging IRLGs, since plenty of scattering material is thought to exist 
in the vicinity of SMBHs.
Hence, the very small scattering efficiency appears not to be natural 
for these sources. 
Future high-quality data at $>$10 keV could be used to see whether 
direct hard X-ray emission is toward us or not, and, in the latter case, 
to estimate the scattering efficiency.

We consider the result suggests that emission in the hard X-ray region 
from SMBH-driven activity in the multinuclei merging IRLGs is 
severely suppressed compared to a typical spectral energy distribution 
of SMBH-activity in AGNs.
For all five radio-excess multinuclei merging IRLGs, 
the presence of strong synchrotron radio emission from SMBH-driven 
activity is suggested.
The presence of strong UV to soft X-ray emission from SMBH-driven 
activity, 
which can account for a predominant fraction of their huge bolometric 
luminosity (Veilleux et al. 1997b; Genzel et al. 1998), 
is also suggested in PKS 1345+12, Mrk 463, and Super-Antennae.
These sources are supposed to have strong SMBH-driven activity,  
perhaps through a high mass accretion rate onto an SMBH, 
and yet hard X-ray luminosity is small.
The hard X-ray emission mechanism in the vicinity of an SMBH is 
thought to be through inverse Compton upscattering of UV to soft X-ray 
photons by high energetic electrons in a hot corona 
(e.g., Walter \& Courvoisier 1992).
It may be that, in merging IRLGs, these high energy electrons are 
not produced as much as in AGNs, or lose energy in a short time scale.
 
In the high-redshift universe, the global (rest-frame) 
far-infrared  emission is dominated by merging IRLGs (Sanders 1999).
Hence, the energy source of merging IRLGs is directly coupled with 
the global interplay between starburst and SMBH-driven activity.
The energy source of these high-redshift merging IRLGs 
will be investigated using the next generation of 
high-sensitivity hard X-ray satellites.
However, if the suppression of hard X-ray emission suggested from 
the study of nearby merging IRLGs is applicable also to 
high-redshift merging IRLGs, without its correction, 
hard X-ray observations would underestimate the contribution of 
SMBH-driven activity in high-redshift merging IRLGs.
We should understand the hard X-ray emission properties from 
SMBH-driven activity in nearby AGNs and merging IRLGs.

\section{Summary}  
       
We found the following main results: 
  
\begin{enumerate}
\item The hard X-ray luminosities relative to far-infrared and 
      infrared  luminosities 
      in the radio-excess multinuclei merging IRLGs are 
      considerably smaller than those in radio-excess AGNs.
\item This result may show that emission in the hard X-ray region 
      from SMBH activity in the multinuclei merging IRLGs is 
      severely suppressed compared to a typical spectral energy 
      distribution of SMBH-driven activity in AGNs.
\item If this is generally applicable to merging IRLGs, 
      without its correction, 
      hard X-ray observations underestimate the contribution of 
      SMBH-driven activity to the bolometric luminosities of merging 
      IRLGs.
\end{enumerate}

\acknowledgments      

We thank all of the members of the {\it ASCA} team who made 
this study possible.
We thank Dr. C. C. Dudley and Ms. L. Good for their careful reading of 
the manuscript, 
Dr. H. Awaki for a useful discussion, and Dr. N. Kobayashi for giving 
us part of his QUIST observing time. 
We are grateful to the anonymous referee for useful comments. 
MI and SU are financially supported by the Japan Society for 
the Promotion of Science for their stays at the University 
of Hawaii and the University of Leicester, respectively.
The Galactic absorption data were obtained from the 
Einstein On-Line 
Service, Smithsonian Astrophysical Observatory (EINLINE).
This research was supported by a Grant-in-Aid 
for Scientific Research on Priority Areas funded by the Ministry of 
Education, Science, Sports, and Culture of Japan.

\clearpage

\figcaption[pks1345_s01.ps]{ 
({\it a}): The SIS X-ray spectrum of PKS 1345+12.
Open circles: SIS0 data. Filled squares: SIS1 data.
The solid and dotted lines are fitted results for the SIS0 and SIS1, 
respectively (see text).
({\it b}): The GIS X-ray spectrum of PKS 1345+12.
Open triangles: GIS2 data. Filled stars: GIS3 data.
The solid and dotted lines are fitted results for the GIS2 and GIS3, 
respectively.
\label{fig1}} 

\figcaption[i19254_s01.ps]{ 
({\it a}) and ({\it b}): The SIS and GIS X-ray spectrum of 
Super-Antennae, respectively.
Same symbols as in Figure 1.
\label{fig2}} 

\figcaption[sgi9259.eps]{ 
{\it Upper}: The {\it q}-value distribution of starburst galaxies at 
high far-infrared luminosities [log($L_{\rm FIR}$/{\it L}$_{\odot}$)  
${\ ^{\displaystyle >}_{\displaystyle =}\ }$ 11.25].
In total, 27 sources are plotted.
No starburst galaxies show {\it q} $<$ 2.05.
We plotted IRLGs with starburst-type and LINER-type optical spectra.
We included IRLGs with LINER-type spectra (11 sources) because shock 
heating driven by superwind activity rather than AGN activity 
is thought to be responsible for their spectra 
(Taniguchi et al. 1999; Lutz, Veilleux, \& Genzel 1999).
Samples are taken from Condon et al. (1991b), and optical classifications 
are from Veilleux et al. (1995).
{\it Lower}: Correlation between $L_{\rm X}$/$L_{\rm FIR}$ and 
observed {\it q}-values of AGNs and multinuclei merging IRLGs. 
Dashed line: {\it q} = 2.0.
We regard sources with {\it q} $<$ 2.0 as radio-excess sources.
Open squares: {\it q} $<$ 2.0 radio-excess AGNs.
Filled circles: {\it q} $<$ 2.0 radio-excess multinuclei merging IRLGs 
(PKS 1345+12, Super-Antennae, IRAS 04154+1755, Mrk 1224, and Mrk 463). 
Open circles: {\it q} $>$ 2.0 AGNs.
For Super-Antennae, IRAS 04154+1755, and Mrk 1224,
we plotted the data by assuming that only a scattered component is seen 
and the scattering efficiency is 3\%.
Note that if direct hard X-ray emission is toward us, the 
$L_{\rm X}$/$L_{\rm FIR}$ is far smaller than the plotted positions.
\label{fig3}}

\figcaption[sgi9259.eps]{ 
{\it Upper}: The {\it q}$'$-value distribution of starburst galaxies at high far-infrared 
luminosities [log($L_{\rm FIR}$/$L_{\odot}$)  
${\ ^{\displaystyle >}_{\displaystyle =}\ }$ 11.25].
The sample sources are the same as in Figure 3.
No starburst galaxies show {\it q}$'$ $<$ 2.1.
{\it Lower}: Correlation between $L_{\rm X}$/$L_{\rm IR}$ and 
observed {\it q}$'$-values. 
Dashed line: {\it q}$'$ = 2.1.
We regard sources with {\it q}$'$ $<$ 2.1 as radio-excess sources.
Open squares: {\it q}$'$ $<$ 2.1 radio-excess AGNs.
Filled circles: {\it q}$'$ $<$ 2.1 radio-excess multinuclei merging 
IRLGs. 
Open circles: {\it q}$'$ $>$ 2.1 AGNs.
\label{fig4}} 

\clearpage
%%%%%%%%%%%%%%%%%%%  Table 1   %%%%%%%%%%%%%%%%%

\scriptsize

\begin{deluxetable}{lccccccrcrc}
%\tablewidth{7.5in}
\tablenum{1}
\tablecaption{Summary of the Properties of the Four Radio-Excess 
Multinuclei Merging IRLGs\label{t}}
\tablecolumns{11}
\tablehead{
\colhead{}&
\colhead{}&
\colhead{$f_{12}$}&
\colhead{$f_{25}$}&
\colhead{$f_{60}$}&
\colhead{$f_{100}$}&
\colhead{$\log L_{\rm FIR}$}&
\multicolumn{1}{c}{}&
\colhead{$\log L_{\rm IR}$}&
\multicolumn{1}{c}{}&
\colhead{}\\
\colhead{Objects}& \colhead{$z$}& \colhead{(Jy)} & \colhead{(Jy)} &
\colhead{(Jy)} &  \colhead{(Jy)} &  \colhead{(ergs s$^{-1})$} &
\multicolumn{1}{c}{$q$}& \colhead{(ergs s$^{-1})$} &
\multicolumn{1}{c}{$q'$}& \colhead{$\frac{f_{25}}{f_{60}}$} \\
\colhead{(1)}& \colhead{(2)}& \colhead{(3)}& \colhead{(4)}& \colhead{(5)}&
\colhead{(6)}&
\colhead{(7)}& \colhead{(8)}& \colhead{(9)}& \colhead{(10)}& \colhead{(11)}}
\startdata
PKS 1345+12     & 0.121 & 0.14 & 0.67 & 1.92 & 2.06 & 45.57
& $-$0.35\tablenotemark{a} & 45.82 & $-$0.10 & 0.35 \\
Super-Antennae & 0.062 & 0.22 & 1.24 & 5.48 & 5.79 & 45.43 &
1.31\tablenotemark{b}& 45.60 & 1.48 &
0.23\\
IRAS 04154+1755 & 0.056 & 0.20 & 0.71 & 3.82 & 5.84 & 45.24 &
1.93\tablenotemark{c} & 45.39 & 2.09 & 0.19 \\
Mrk 1224 & 0.050 & 0.18 & 0.50 & 4.12 & 6.98 & 45.19 & 
1.97\tablenotemark{c}
& 45.30 & 2.08 & 0.12 \\
%\tableline
\enddata
%\tablenotetext{a}{Calculated with
%            $L_{\rm FIR} = 2.1 \times 10^{39} \times$ D(Mpc)$^{2}$
%            $\times (2.58 \times f_{60} + f_{100}$).}
%\tablenotetext{a}{Calculated with
%            $L_{\rm IR} = 2.1 \times 10^{38} \times$ D(Mpc)$^{2}$
%            $\times$ (13.48 $\times$ $f_{12}$ + 5.16 $\times$ $f_{25}$ +
%            $2.58 \times f_{60} + f_{100}$).}

\tablenotetext{a}{Kim 1995.}
\tablenotetext{b}{Roy \& Norris 1997.}
\tablenotetext{c}{Crawford et al. 1996.} 

\normalsize

\tablecomments{Column (1): Object name. Column (2): Redshift.
Column (3): IRAS 12 $\mu$m flux in Jy.
Column (4): IRAS 25 $\mu$m flux in Jy.
Column (5): IRAS 60 $\mu$m flux in Jy.
Column (6): IRAS 100 $\mu$m flux in Jy.
Column (7): Logarithm of far-infrared (40$-$500$\mu$m) luminosity 
            in ergs s$^{-1}$ calculated with
            $L_{\rm FIR} = 2.1 \times 10^{39} \times$ D(Mpc)$^{2}$
            $\times (2.58 \times f_{60} + f_{100}$) [ergs s$^{-1}$]
            (Sanders \& Mirabel 1996).
Column (8): Observed $q$-value.
Column (9): Logarithm of infrared (8$-$1000 $\mu$m) luminosity 
            in ergs s$^{-1}$ calculated with
            $L_{\rm IR} = 2.1 \times 10^{39} \times$ D(Mpc)$^{2}$
            $\times$ (13.48 $\times$ $f_{12}$ + 5.16 $\times$ $f_{25}$ +
            $2.58 \times f_{60} + f_{100}$) [ergs s$^{-1}$]
            (Sanders and Mirabel 1996).
Column (10): Observed $q'$-value
             [$\equiv  \log(f_{\rm IR}/f_{\rm 20 cm}$)].
Column (11): IRAS 25 $\mu$m to 60 $\mu$m flux ratio.}

\end{deluxetable}

\clearpage 
%%%%%%%%%%%%%%%%%  Table 2 %%%%%%%%%%%%%%%%%

\normalsize

\begin{deluxetable}{llcccc}
%\tablewidth{7.5in}
\tablenum{2}
\tablecaption{Observing Log\label{h}}
\tablecolumns{6}
\tablehead{
\colhead{} & \colhead{}& \multicolumn{2}{c}{Integration\tablenotemark{a}} &
\multicolumn{2}{c}{Net Counts\tablenotemark{b}} \nl
\cline{3-4} \cline{5-6}\nl
\colhead{Objects} & \colhead{Observing Date} & \colhead{SIS} & \colhead{GIS} &
\colhead{SIS} & \colhead{GIS}}
\startdata
PKS 1345+12     & 1996 Jul 20   & 19 & 20 & $309 \pm 21$ & $311 \pm 16$\nl
Super-Antennae & 1996 Oct 16  & 33 & 37 & $191 \pm 21$ & $204 \pm 21$\nl
IRAS 04154+1755 & 1998 Feb 15  &31 & 35 & $82 \pm 18$ & $15 \pm 20$\nl
Mrk 1224        & 1998 Nov 11 & 21 & 24 & $72 \pm 14$ & $63 \pm 17$ \nl
\enddata
\tablenotetext{a}{The SIS and GIS exposure time in ksec, respectively.}
\tablenotetext{b}{Net source counts of the sum of SIS0 + SIS1
and GIS2 + GIS3, respectively.  Errors are background photon noise (1
$\sigma$).}

%\tablecomments{Column (1): Object name. Column (2): Observing date.
%Columns (3) and (4): The SIS and GIS exposure time in ksec,
%respectively.
%Columns (5) and (6): Net source counts of the sum of SIS0 + SIS1
%and GIS2 + GIS3, respectively.
%Errors are background photon noise (1 $\sigma$).}

\end{deluxetable}

%\end{document}

\clearpage
%%%%%%%%%%%%%%%%%  Table 3 %%%%%%%%%%%%%%%%%

\scriptsize

\begin{deluxetable}{lccccc}
%\tablewidth{7.5in}
\tablenum{3}
\tablecaption{Fitting Results of the Four Radio-Excess Multinuclei 
Merging IRLGs}
\tablecolumns{6}
\tablehead{
\colhead{}& \colhead{}& \colhead{$N_{\rm H}$}&
\colhead{$f_{\rm X}$(2--10 keV)}&
\colhead{$L_{\rm X}$(2--10 keV)}&
\colhead{$\chi^{2}$/dof} \\
\colhead{Objects}&  \colhead{$\Gamma$} & \colhead{(10$^{21}$ cm$^{-2}$)} &
\colhead{(ergs s$^{-1}$ cm$^{-2}$)}  & \colhead{(ergs s$^{-1}$)} & \\
\colhead{(1)}& \colhead{(2)}& \colhead{(3)}& \colhead{(4)}& \colhead{(5)}&
\colhead{(6)}}
\startdata
%%%%%%%%%%%%%%%%%  PKS 1345+12 %%%%%%%%%%%%%%%%%%%%%%%%%%%%%%%%%%%%%
PKS 1345+12 & 0.82$^{+0.07}_{-0.03}$ & 15$^{+4}_{-3}$ &
1.0 $\times$ 10$^{-12}$ & 3.2 $\times$ 10$^{43}$ & 55.0/51 \\
            & 1.7 (fixed) & 39$^{+6}_{-3}$ &
8.2 $\times$ 10$^{-13}$ & 3.0 $\times$ 10$^{43}$ & 61.1/52 \\
\tablevspace{2pt}\tableline\tablevspace{2pt}
%%%%%%%%%%%%%%%%%  Super-Antennae %%%%%%%%%%%%%%%%%%%%%%%%%%%%%%%%%%%
Super-Antennae & 1.1$^{+0.1}_{-0.2}$ & $<$0.2 &
1.9 $\times$ 10$^{-13}$ & 1.4 $\times$ 10$^{42}$ & 35.3/51 \\
                & 1.7 (fixed) & 3.1$^{+3.9}_{-0.7}$ &
1.5 $\times$ 10$^{-13}$ & 1.2 $\times$ 10$^{42}$ & 38.0/52 \\ 
                & 1.7 (fixed) & $>$1000 (fixed) & 
1.1 $\times$ 10$^{-13}$ 
& 2.9 $\times$ 10$^{43}$ $\times$ (0.03/{\it f}$_{\rm scatt}$) 
& 39.1/53 \\ \hline
%%%%%%%%%%%%%%%%%  I04154 %%%%%%%%%%%%%%%%%%%%%%%%%%%%%%%%%%%%%%%%%%%%
IRAS 04154+1755 & 1.0 (fixed) & 0.0 (fixed) & 
$<$9.6 $\times$ 10$^{-14}$ & 
$<$5.9 $\times$ 10$^{41}$ & --- \\
                & 1.7 (fixed) & 100 (fixed) & $<$9.1 $\times$ 10$^{-14}$ 
 & $<$9.2 $\times$ 10$^{41}$ & --- \\ 
                & 1.7 (fixed) & $>$1000 (fixed) & 
$<$5.5 $\times$ 10$^{-14}$ 
& $<$1.1 $\times$ 10$^{43}$ $\times$ (0.03/{\it f}$_{\rm scatt}$) 
& --- \\ \hline
%%%%%%%%%%%%%%%%%  Mrk 1224 %%%%%%%%%%%%%%%%%%%%%%%%%%%%%%%%%%%%%%%%%
Mrk 1224 & 1.0 (fixed) & 0.0 (fixed) & $<$1.7 $\times$ 10$^{-13}$ 
& $<$8.4 $\times$ 10$^{41}$ & --- \\
         & 1.7 (fixed) & 100 (fixed) & $<$1.5 $\times$ 10$^{-13}$ 
& $<$1.2 $\times$ 10$^{42}$ & --- \\
         & 1.7 (fixed) & $>$1000 (fixed) & $<$1.0 $\times$ 10$^{-13}$ 
& $<$1.7 $\times$ 10$^{43}$ $\times$ (0.03/{\it f}$_{\rm scatt}$) & --- \\
%%%%%%%%%%%%%%%%%%%%%%%%%%%%%%%%%%%%%%%%%%%%%%%%%%%%%%%%%%%%%%%%%%%%%
\enddata

\vspace{1cm}

\tablecomments{
Column (1): Object name.
Column (2): Power-law photon index. Errors are for the 1 $\sigma$ level
($\Delta \chi^{2} = 1.0$ for one interesting parameter).
Column (3): Column density of absorbing gas at the rest frame.
            Errors are for the 1 $\sigma$ level.
Column (4): Hard X-ray flux. 
Column (5): Extinction-corrected hard X-ray luminosity.
Column (6): Reduced $\chi^{2}$ value.
}
\end{deluxetable}

\clearpage
%%%%%%%%%%%% Table 4  %%%%%%%%%%%%%%
\begin{deluxetable}{lccc}
%\tablewidth{7.5in}
\tablenum{4}
\tablecaption{Summary of the Hard X-ray Properties of the Four 
Radio-Excess Multinuclei Merging IRLGs}
\tablecolumns{4}
\tablehead{
\colhead{}& \multicolumn{1}{c}{$\log L_{\rm X}$} & \colhead{}& \colhead{}\\
\colhead{Objects} &  \multicolumn{1}{c}{(ergs s$^{-1}$)}&
\multicolumn{1}{c}{$\log \frac{L_{\rm X}}{L_{\rm FIR}}$} &
\multicolumn{1}{c}{$\log \frac{L_{\rm X}}{L_{\rm IR}}$ } \\
\colhead{(1)}& \colhead{(2)}& \colhead{(3)}& \colhead{(4)}
}
\startdata
PKS 1345+12 & 43.50 & $-$2.07 & $-$2.32 \\
Super-Antennae & 42.15 (43.46) & $-$3.28 ($-$1.97) & $-$3.45 ($-2.14$) \\
IRAS 04154+1755 & $<$41.96 ($<$43.04) & $<$ $-$3.28 ($<$ $-$2.20) 
& $<$ $-$3.43 ($<$ $-$2.35)\\
Mrk 1224 &$<$42.08 ($<$43.23) & $<$ $-$3.11 ($<$ $-$1.96) 
& $<$ $-$3.22 ($<$ $-$2.07) \\
\enddata
\tablecomments{Column (1): Object name.
Column (2): Logarithm of extinction-corrected hard X-ray luminosity.
%            in ergs s$^{-1}$.
Column (3): Logarithm of hard X-ray to  far-infrared luminosity ratio.
Column (4): Logarithm of hard X-ray to infrared  luminosity ratio.
Values in parentheses are for the case when 
only a scattered component is seen and the scattering efficiency 
is 3\%.}

\end{deluxetable}

\normalsize

\clearpage

%%%%%%%%%%%%% Table 5 %%%%%%%%%%%%%
%\documentstyle[apjpt4]{article}
%\documentstyle[12pt,aasms4]{article}
%\begin{document}
\begin{deluxetable}{lcccrccrrcc}
\tablewidth{7.06in}
\tablenum{5}
\tablecaption{Summary of the Properties of AGNs and Radio-Excess 
Merging IRLGs in the Literature}
\scriptsize
\tablecolumns{11}
%\tablehead{
%\colhead{Objects} & \colhead{$z$} & \colhead{$\log L_{\rm X}$} &
%\colhead{$\log L_{\rm FIR}$} &
%\multicolumn{1}{c}{$\log \frac{L_{\rm X}}{L_{\rm FIR}}$} &\colhead{$q$} &
%\colhead{$\log L_{\rm IR}$} &
%\multicolumn{1}{c}{$\log \frac{L_{\rm X}}{L_{\rm IR}}$} &
%\colhead{$q'$} &
%\colhead{$\frac{f_{25}}{f_{60}}$} & \colhead{class}}
\tablehead{
\colhead{}&\colhead{}& \colhead{$\log L_{\rm X}$} 
& \colhead{$\log L_{\rm FIR}$}  &
\colhead{}&\colhead{}&
\colhead{$\log L_{\rm IR}$} &
\colhead{}&\colhead{}& \colhead{}& \colhead{} \\
\colhead{Objects} & \colhead{$z$} &  \colhead{(ergs s$^{-1}$)}&
\colhead{(ergs s$^{-1}$)}&
\multicolumn{1}{c}{$\log \frac{L_{\rm X}}{L_{\rm FIR}}$} &
\colhead{$q$} &
\colhead{(ergs s$^{-1}$)}&
\multicolumn{1}{c}{$\log \frac{L_{\rm X}}{L_{\rm IR}}$} &
\multicolumn{1}{c}{$q'$} &
\colhead{$\frac{f_{25}}{f_{60}}$}  & \colhead{Class}\\
\colhead{(1)}& \colhead{(2)}& \colhead{(3)}& \colhead{(4)}&
\multicolumn{1}{c}{(5)}&
\colhead{(6)}&
\colhead{(7)}& \multicolumn{1}{c}{(8)}& \multicolumn{1}{c}{(9)}&
\colhead{(10)}&
\colhead{(11)}}
\startdata
\multicolumn{11}{c}{$q < 2.0$ radio-excess multinuclei merging IRLGs} \\ 
\tableline
Mrk 463  & 0.050 & 42.40$^{a}$ & 44.82 & $-$2.42 & 0.83$^{j}$ & 45.32 & 
$-$2.92 &
1.33 & 0.74 & N \\
\cutinhead{$q < 2.0$ radio-excess AGNs}
Mrk 937 & 0.030 & 42.53$^{a}$ & 43.91 & $-$1.38 & $>$1.60$^{k,1}$ &
44.26 & $-$1.73 & $>$1.95 & 0.46 & N \\
NGC 526A & 0.019 & 43.42$^{a}$ (42.83$^{b}$) & 43.26 & 0.16 & 1.67$^{l}$
& 43.99 & $-$0.57 & 2.40 & 2.0 & N \\
NGC 2110 & 0.008 & 42.47$^{a}$ (42.50$^{b}$) & 43.55 & $-$1.08
& 1.32$^{l}$ & 43.73 & $-$1.26 & 1.50 & 0.20 & N \\
NGC 2992 & 0.008 & 41.70$^{a}$ (42.31$^{b}$)& 43.82 & $-$2.12
& 1.71$^{m}$ & 43.98 & $-$2.28 & 1.87 & 0.20 & N \\
NGC 4968 & 0.010 & 42.10$^{a}$ & 43.51 & $-$1.41 & 1.93$^{j}$ &
43.86 & $-$1.76 & 2.28 & 0.53 & N \\
IC 5063 & 0.011 & 42.80$^{a}$ & 43.94 & $-$1.14 & 0.76$^{n,2}$
& 44.39 & $-$1.59 & 1.21 & 0.64 & N \\
Mrk 335 & 0.025 & 43.07$^{c}$ (43.24$^{b}$)& 43.50 & $-$0.43 &
1.86$^{l}$ & 44.21 & $-$1.14 & 2.57 & 1.1\phn & B  \\
3C 120 & 0.033 & 43.99$^{c}$ & 44.36 & $-$0.37 & $-$0.33$^{j}$ &
44.73 & $-$0.74 & 0.04 & 0.54 & B \\
IC 4329A & 0.016 & 43.59$^{c}$ (43.83$^{b}$) & 43.78 & $-$0.19
& 1.59$^{l}$ & 44.46 & $-$0.87 & 2.27 & 1.1\phn & B \\
NGC 5548 & 0.017 & 43.41$^{c}$ (43.31$^{b}$) & 43.64 & $-$0.23
& 1.81$^{l}$ & 44.12 & $-$0.71 & 2.29 & 0.73 & B \\
NGC 5506 & 0.006 & 42.73$^{b}$ & 43.60 & $-$0.87 & 1.60$^{j}$ &
43.93 & $-$1.20 & 1.93 & 0.41 & N \\
3C 390.3 & 0.056 & 44.46$^{b}$ & 44.10 & 0.36 & $-$0.54$^{o,3}$ &
44.68 & $-$0.22 & 0.04 & 1.4\phn & B \\
NGC 7213 & 0.006 & 42.43$^{b}$ & 43.27 & $-$0.84 & 1.48$^{n,4}$ &
43.53 & $-$1.10 & 1.74 & 0.29 & B \\
Q0637-752 & 0.651 & 45.79$^{d}$& $<$46.22\phantom{$<$} & $>$ $-$0.43 &
$<$ $-$1.21$^{d,3}$ & $<$46.67\phantom{$<$} & $>$ $-$0.88 & $<$ $-$0.76 & 0.65
& B \\
Q0837-120 & 0.198 & 44.69$^{d}$& $<$44.53\phantom{$<$} & $>$0.16 &
$<$ $-$0.29$^{d,3}$ & $<$45.12\phantom{$<$} & $>$ $-$0.43 & $<$0.30 &
$<$0.70\phantom{$<$} & B
\\ Mrk 279 & 0.031 & 43.38$^{d}$ & 44.23 & $-$0.85 & 1.88$^{d}$ &
$<$44.52\phantom{$<$} & $>$ $-$1.14 & $<$2.17 & 0.24 & B \\
Q2135$-$147 & 0.200 & 44.91$^{d}$& $<$45.16\phantom{$<$} & $>$ $-$0.25 &
$<$0.57$^{d}$ & $<$45.68\phantom{$<$} & $>$ $-$0.77 & $<$1.09 & 0.82 & B \\
Mrk 3 & 0.014 & 42.66$^{e,5}$ (43.33$^{e}$) & 43.93 & $-$1.27
& 0.63$^{l}$ & 44.38 & $-$1.72 & 1.08 & 0.73 &  N \\
Mrk 348 & 0.015 & 43.02$^{f,6}$ & 43.63 & $-$0.61 & 0.79$^{l}$
& 44.03 & $-$1.01 & 1.19 & 0.53 & N \\
\cutinhead{$q > 2.0$ AGNs}
NGC 1667 & 0.015 & 40.11$^{a}$ & 44.37 & $-$4.26 & 2.32$^{p}$ &
44.49 & $-$4.38 & 2.44 & 0.11 & N \\
NGC 1808 & 0.003 & 40.27$^{a}$ & 44.16 & $-$3.89 & 2.40$^{q}$ &
44.29 & $-$4.02 & 2.53 & 0.16 & N \\
NGC 4507 & 0.012 & 41.93$^{a}$ & 43.92 & $-$1.99 & 2.21$^{n,2}$
& 44.17 & $-$2.24 & 2.46 & 0.31 & N \\
NGC 5135 & 0.014 & 42.81$^{a}$ & 44.67 & $-$1.86 & 2.10$^{j}$ &
44.79 & $-$1.98 & 2.22 & 0.15 & N \\
ESO 103$-$G35 & 0.013 & 42.79$^{a}$ & 43.63 & $-$0.84 & 2.29$^{r,2}$ &
44.22 & $-$1.43 & 2.88 & 1.04 & N \\
NGC 7172 & 0.009 & 42.87$^{a}$ (42.69$^{b}$) & 43.85 & $-$0.98 &
2.50$^{j}$ & 43.99 & $-$1.12 & 2.64 & 0.13 & N \\
NGC 7314 & 0.005 & 42.24$^{a}$ (42.29$^{b}$) & 43.25 & $-$1.01 &
2.37$^{q}$ & 43.36 & $-$1.12 & 2.48 & 0.15 & N \\
NGC 7582 & 0.005 & 41.97$^{a}$ & 44.27 & $-$2.30 & 2.39$^{q}$ & 44.38
& $-$2.41 & 2.50 & 0.13 & N \\
NGC 3227 & 0.003 & 41.66$^{c}$ (42.10$^{b}$) & 43.07 & $-$1.41
& 2.10$^{j}$ & 43.24 & $-$1.58 & 2.27 & 0.22 & B  \tablebreak
NGC 3516 & 0.009 & 43.08$^{c}$ (42.54$^{b}$) & 43.26 & $-$0.18
& 2.38$^{l}$ & 43.69 & $-$0.61 & 2.81 & 0.53 & B \\
NGC 3783 & 0.009 & 42.90$^{c}$ (43.05$^{b}$) & 43.58 & $-$0.68 &
$>$2.78$^{s,2}$ & 44.01 & $-$1.11 & $>$3.21 & 0.72 & B \\
NGC 4051 & 0.002 & 41.21$^{c}$ (41.13$^{b}$) & 42.76 & $-$1.55
& 2.54$^{j}$ & 42.92 & $-$1.71 & 2.70 & 0.17 & B \\
Mrk 766 & 0.012 & 42.73$^{c}$ & 43.86 & $-$1.13 & 2.10$^{l}$
& 44.13 & $-$1.40 & 2.37 & 0.34 & B \\
NGC 4593 & 0.009 & 42.71$^{c}$ (42.69$^{b}$) & 43.56 & $-$0.85 &
3.33$^{t}$ & 43.80 & $-$1.09 & 3.57 & 0.33 & B \\
MCG$-$6-30-15 & 0.008 & 42.72$^{c}$ (42.83$^{b}$) & 42.93 & $-$0.21
& 2.89$^{m}$ & 43.46 & $-$0.74 & 3.42 & 0.74 & B \\
Mrk 841 & 0.036 & 43.47$^{c}$ (43.33$^{b}$) & 43.91 & $-$0.44 &
$>$2.00$^{m}$ & 44.49 & $-$1.02 & $>$2.58 & 1.0\phn & B \\
NGC 6814 & 0.006 & 40.93$^{c}$ & 43.61 & $-$2.68 & 3.27$^{m}$ &
43.70 & $-$2.77 & 3.36 & 0.10 & B \\
Mrk 509  & 0.035 & 44.03$^{c}$ (44.10$^{b}$) & 44.34 & $-$0.31 &
2.12$^{m}$ & 44.76 & $-$0.73 & 2.54 & 0.52 & B \\
NGC 7469 & 0.017 & 43.25$^{c}$ (43.31$^{b}$)& 45.02 & $-$1.77
& 2.31$^{j}$ & 45.18 & $-$1.93 & 2.47 & 0.20 & B \\
Akn 120 & 0.033 & 43.88$^{b}$ & 44.01 & $-$0.13 & 2.06$^{u}$ &
44.54 & $-$0.66 & 2.59 & 0.64 & B \\
Mrk 705 & 0.028 & 43.25$^{d}$ & 43.87 & $-$0.62 & $>$2.69$^{d}$ &
44.24 & $-$0.99 & $>$3.06 & 0.47 & B \\
Q2130+099 & 0.061 & 43.57$^{d}$& 44.28 & $-$0.71 & 2.02$^{d}$ &
44.85 & $-$1.28 & 2.59 & 0.66 & B \\
Mrk 231 & 0.042 & 42.34$^{g}$ & 45.86  & $-$3.52 & 2.22$^{v}$
& 46.06 & $-$3.72 & 2.42 & 0.25 & B/U \\
IRAS F20460+1925 & 0.181 & 44.08$^{h}$ & 45.52 & $-$1.44 & 2.18$^{r,2}$
& 46.02 & $-$1.94 & 2.68 & 0.65 & N/U \\
IRAS F23060+0505 & 0.174 & 44.18$^{i}$ & 45.63 & $-$1.45 &
$>$2.62$^{r,2}$ & 45.99 & $-$1.81 & $>$2.98 & 0.37 & N/U \\
\enddata

%\normalsize
\small

\tablecomments{
Column (1): Object name. Column (2): Redshift.
Column (3): Logarithm of extinction-corrected hard X-ray luminosity.
            If both the {\it ASCA} and {\it Ginga} data are available,
            we have adopted {\it ASCA} data and showed {\it Ginga} 
            data for comparison in parentheses.
Column (4): Logarithm of far-infrared luminosity.
Column (5): Logarithm of hard X-ray to far-infrared luminosity ratio.
Column (6): Observed $q$-value.  
            For radio-loud AGNs with extended radio emission, we have
            adopted the core radio flux.
Column (7): Logarithm of infrared luminosity.
Column (8): Logarithm of hard X-ray to infrared luminosity ratio.
Column (9): Observed $q'$-value. 
            For radio-loud AGNs with extended radio emission, we have
            adopted the core radio flux.
Column (10): IRAS 25 $\mu$m to 60 $\mu$m flux ratio.
Column (11): Classification.
B = broad optical line AGN; N = narrow optical line AGN; 
U = ultraluminous infrared  galaxy,
$L_{\rm IR} >10^{12}L_{\odot}$ (Sanders \& Mirabel 1996).}
\end{deluxetable}

\clearpage

%\normalsize
\small

$^{1}$: 1.5 GHz radio flux was estimated from 4.8 GHz radio flux
         by assuming a radio spectrum of $\nu^{-1}$.

$^{2}$: 1.5 GHz radio flux was estimated from 2.3 GHz radio flux
         by assuming a radio spectrum of $\nu^{-1}$.

$^{3}$: 1.5 GHz radio flux was estimated from 1.4 GHz radio flux
         by assuming a radio spectrum of $\nu^{-1}$.

$^{4}$: 1.5 GHz radio flux was estimated from 2.3 GHz radio flux
        by assuming a radio spectrum of $\nu^{0}$, since this object
        shows a flat radio spectrum between 0.8 GHz and 5 GHz
        (Bransford et al. 1998; Harnett 1987).

$^{5}$: {\it ASCA} data indicate that the detected hard X-ray emission 
        is a scattered component (Netzer et al. 1998),
        but Griffiths et al. (1998) estimated the hard X-ray
        luminosities for both the {\it Ginga} and {\it ASCA} data 
        by analyzing the combined {\it ASCA}, {\it Ginga}, and 
        {\it ROSAT} data.

$^{6}$: Netzer et al. (1998) considered the detected hard X-ray
        spectrum is a scattered component based on its spectral shape.
        However, the equivalent width of the iron K$\alpha$ emission 
        line is small ($\sim$100 eV), so we regard this source as 
        the one whose direct hard X-ray emission is detected.
        The $L_{\rm X}$ based on {\it ASCA} data is not available, so
        we adopted {\it Ginga} data.

References. --- $^{a}$ Turner et al. 1997; $^{b}$ Nandra \& Pounds 1994; 
$^{c}$ Nandra et al. 1997; $^{d}$ Elvis et al. 1994;
$^{e}$ Griffiths et al. 1998; $^{f}$ Alonso-Herrero, Ward, \& Kotilainen 1997;
$^{g}$ Iwasawa 1999; $^{h}$ Ogasaka et al. 1997;
$^{i}$ Brandt et al. 1997; $^{j}$ Rush, Malkan, \& Edelson 1996; 
$^{k}$ Bicay et al. 1995; $^{l}$ Nagar et al. 1999; 
$^{m}$ Dahari \& De Robertis 1988;
$^{n}$ Bransford et al. 1998; $^{o}$ Wall \& Peacock 1985;
$^{p}$ Condon et al. 1990; $^{q}$ Condon 1987; 
$^{r}$ Heisler et al. 1998; $^{s}$ Roy et al. 1994; 
$^{t}$ Ulvestad \& Wilson 1984; 
$^{u}$ Barvainis, Lonsdale, \& Antonucci 1996; 
$^{v}$ Condon et al. 1991b.

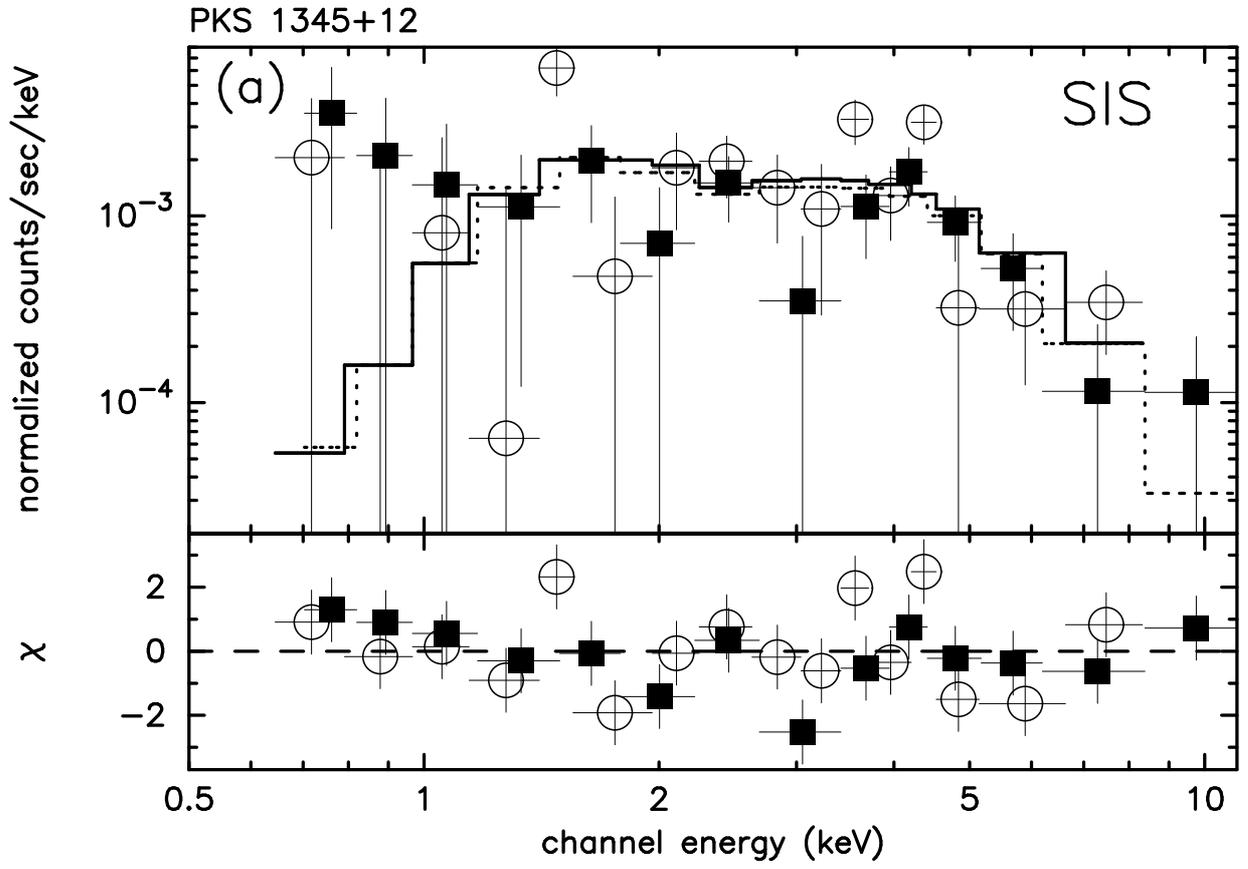
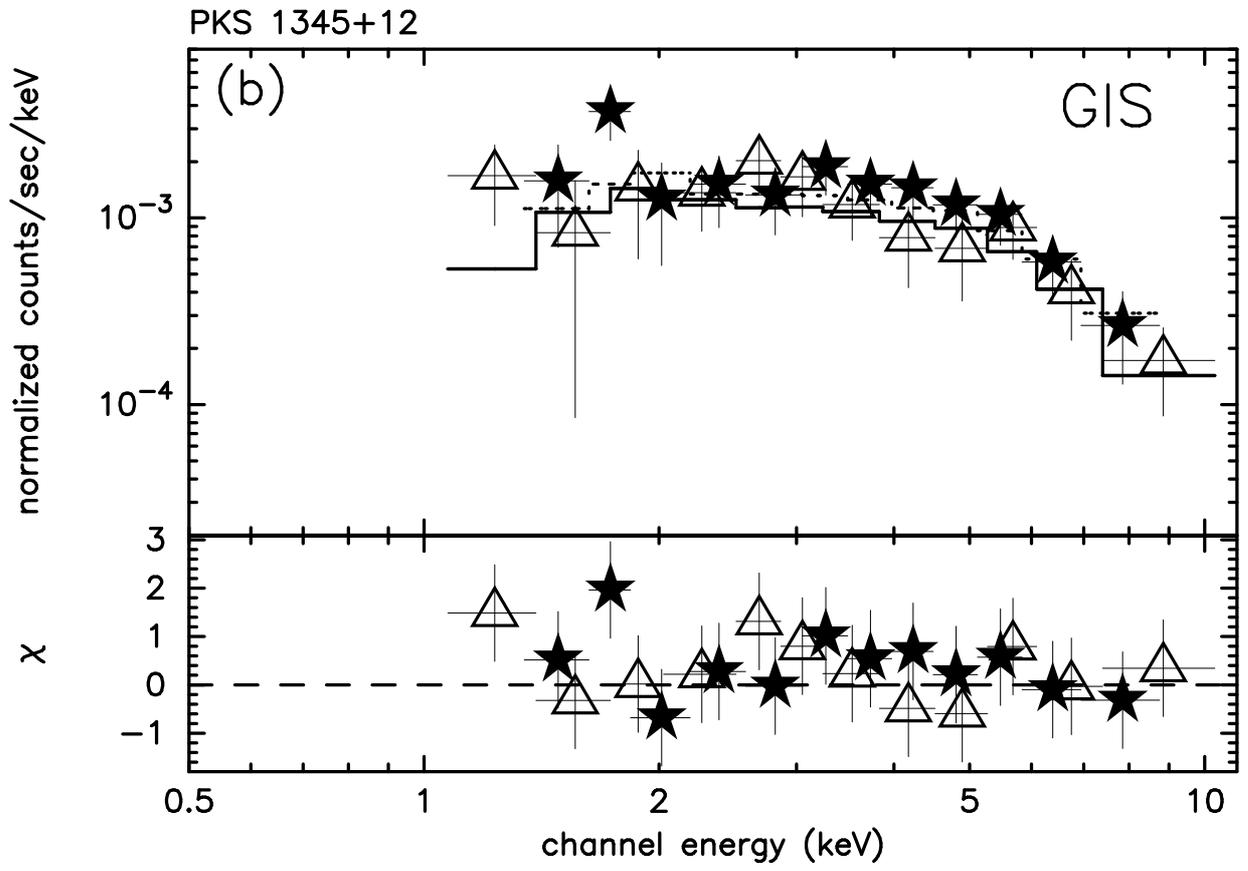
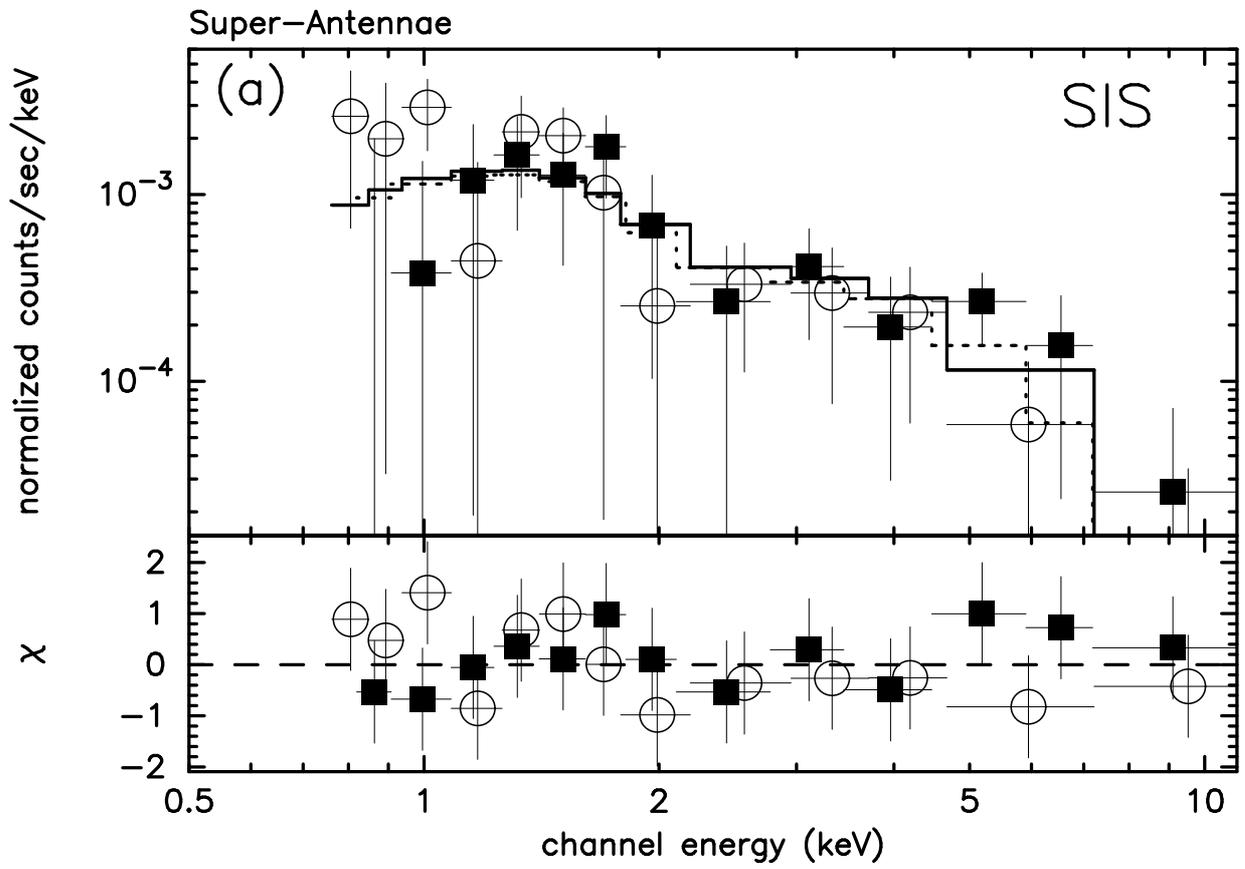
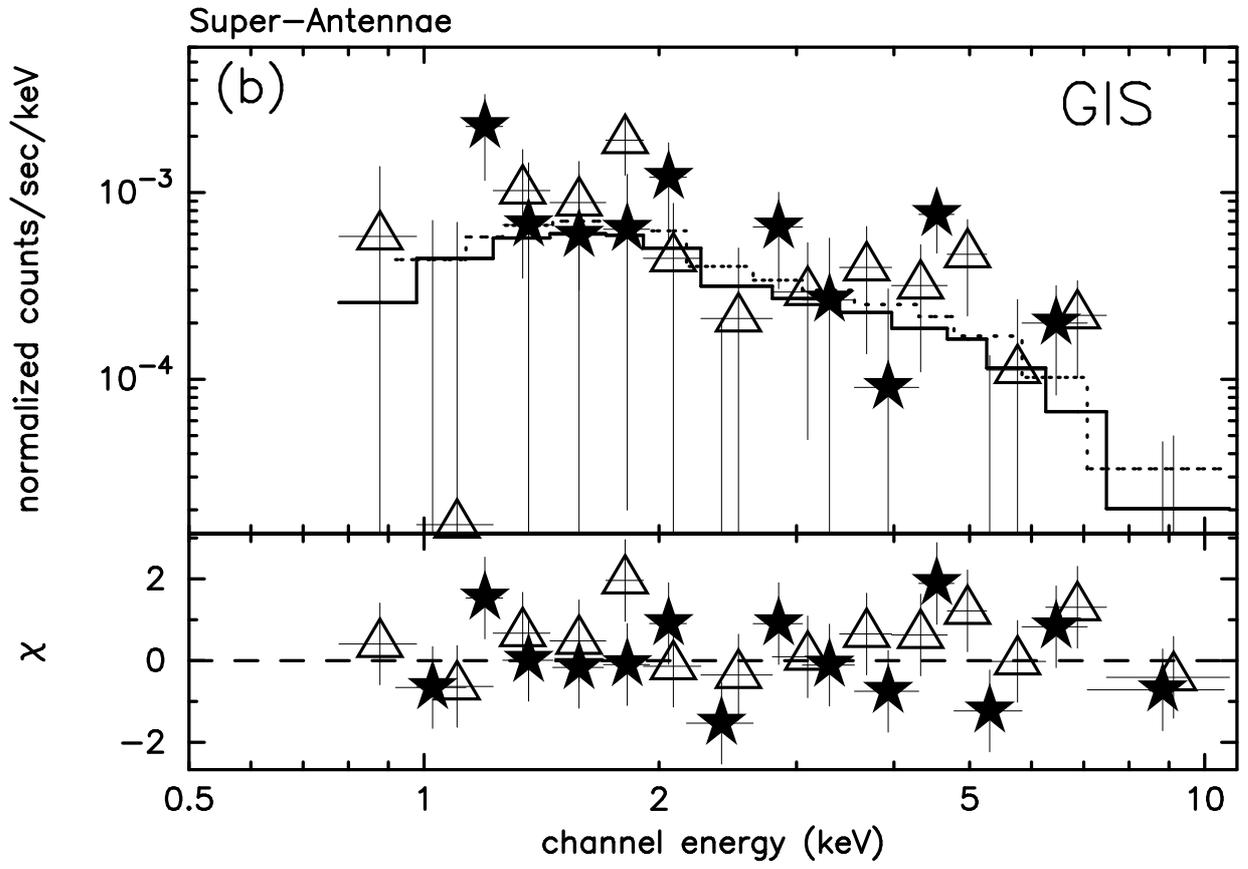
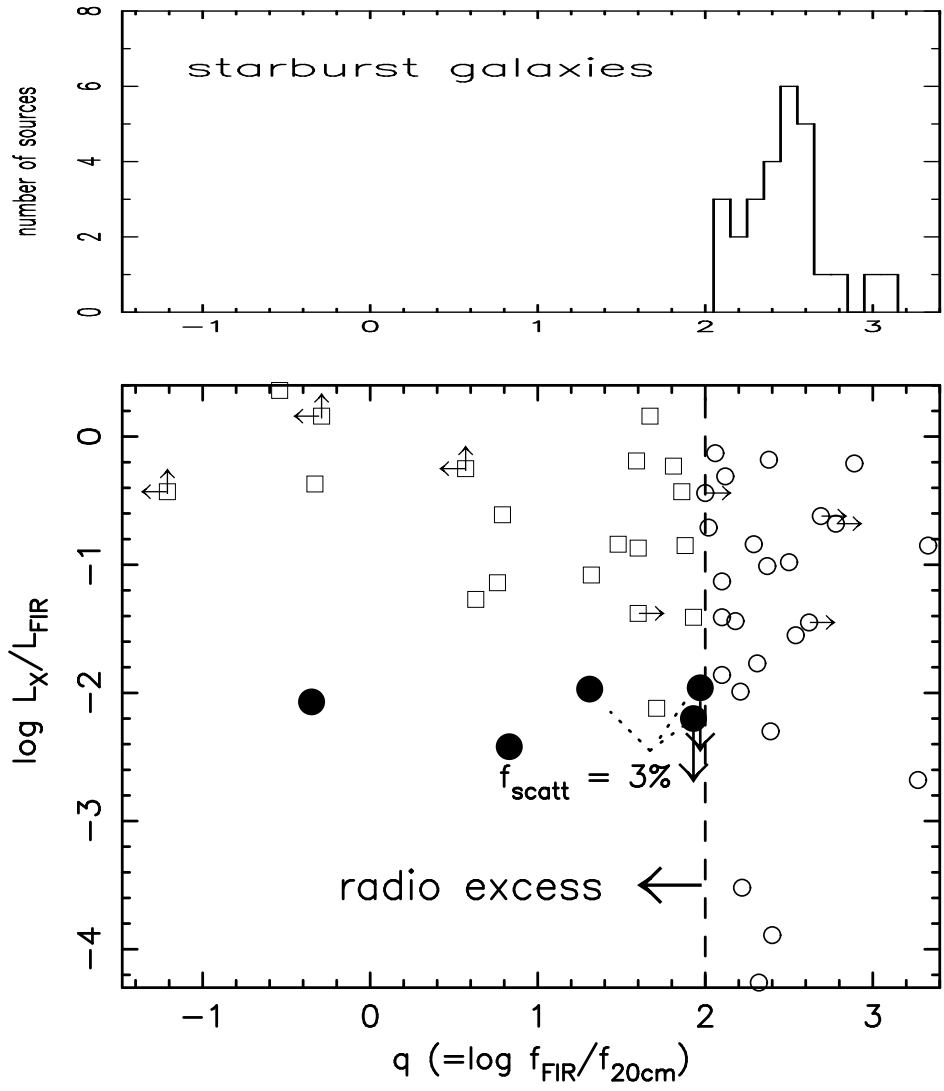
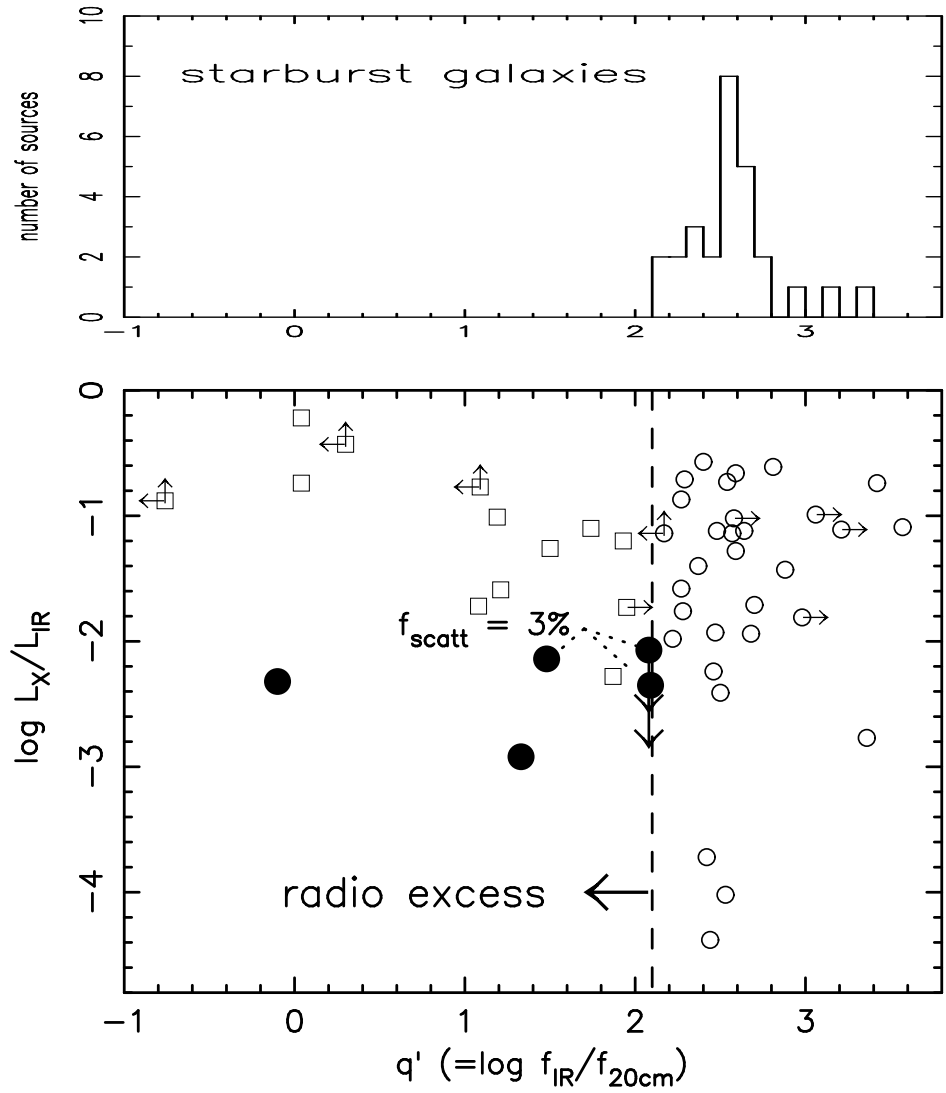
\end{document}